\documentclass[%
 amsmath,amssymb,amsfonts,
 aps,
twocolumn
]{revtex4-2}

\usepackage [english]{babel}
\usepackage [autostyle, english = american]{csquotes}
\MakeOuterQuote{"}

\usepackage{appendix}

\usepackage[normalem]{ulem}

\usepackage{graphicx}% Include figure files
\usepackage{dcolumn}% Align table columns on decimal point
\usepackage{bm}% bold math
\usepackage{hyperref}% add hypertext capabilities
\usepackage[mathlines]{lineno}% Enable numbering of text and display math
\usepackage[%showframe,%Uncomment any one of the following lines to test 
margin=1in,
]{geometry}

\usepackage[export]{adjustbox}

\usepackage{pgfplots}
\usetikzlibrary{arrows.meta}
\usetikzlibrary{backgrounds}
\usepgfplotslibrary{patchplots}
\usepgfplotslibrary{fillbetween}
\pgfplotsset{%
    layers/standard/.define layer set={%
        background,axis background,axis grid,axis ticks,axis lines,axis tick labels,pre main,main,axis descriptions,axis foreground%
    }{
        grid style={/pgfplots/on layer=axis grid},%
         tick style={/pgfplots/on layer=axis ticks},%
         axis line style={/pgfplots/on layer=axis lines},%
         label style={/pgfplots/on layer=axis descriptions},%
         legend style={/pgfplots/on layer=axis descriptions},%
         title style={/pgfplots/on layer=axis descriptions},%
         colorbar style={/pgfplots/on layer=axis descriptions},%
         ticklabel style={/pgfplots/on layer=axis tick labels},%
         axis background@ style={/pgfplots/on layer=axis background},%
         3d box foreground style={/pgfplots/on layer=axis foreground},%
     },
 }

\usepackage[normalem]{ulem} 

\usepackage{physics}

\usepackage{caption}
\usepackage{subcaption}

\makeatletter
\newcommand*{\rightharpoonupfill@}{%
  \arrowfill@\relbar\relbar\rightharpoonup
}

\newcommand*{\leftharpoondownfill@}{%
  \arrowfill@\leftharpoondown\relbar\relbar
}

\newcommand{\xrightleftharpoons}[2][]{%
  \ensuremath{%
    \mathrel{%
      \settoheight{\dimen@}{\raise 2pt\hbox{$\rightharpoonup$}}%
      \setlength{\dimen@}{-\dimen@}%
      \edef\CA@temp{\the\dimen@}%
      \settoheight\dimen@{$\rightleftharpoons$}%
      \addtolength{\dimen@}{\CA@temp}%
      \raisebox{\dimen@}{%
        \rlap{%
          \raisebox{2pt}{%
            $%
            \ext@arrow 0359\rightharpoonupfill@{\hphantom{#1}}{#2}%
            $%
          }%
        }%
        \hbox{%
          $%
          \ext@arrow 3095\leftharpoondownfill@{#1}{\hphantom{#2}}%
          $%
        }%
      }%
    }%
  }%
}
\makeatother

\usepackage{soul}

\begin{document}

\title{Dynamics of qudit gates and effects of spectator modes on \\optimal control pulses} 

\author{A. Bar\i\c{s} \"Ozg\"uler}
\email{baris\_ozguler@berkeley.edu}
\affiliation{%
  Fermi National Accelerator Laboratory, Batavia, IL, 60510
}%
\affiliation{%
  Superconducting Quantum Materials and Systems Center (SQMS), Fermilab
}%

\author{
Joshua A. Job
}
\affiliation{Lockheed Martin Advanced Technology Center, Sunnyvale, CA, 94089}
\affiliation{%
  Superconducting Quantum Materials and Systems Center (SQMS), Fermilab
}%

\date{March 25, 2024}

\begin{abstract}

Qudit gates for high-dimensional quantum computing can be synthesized with high precision using numerical quantum optimal control techniques. Large circuits are broken down into modules and the tailored pulses for each module can be used as primitives for a  qudit compiler. Application of the pulses of each module in the presence of extra modes may decrease their effectiveness due to crosstalk. In this paper, we address this problem by simulating qudit dynamics for circuit quantum electrodynamics (cQED) systems. As a test case, we take pulses for single-qudit SWAP gates optimized in isolation and then apply them in the presence of spectator modes each of which are in Fock states. We provide an experimentally relevant scaling formula that can be used as a bound on the fidelity decay. Our results show that frequency shift from spectator mode populations has to be $\lesssim 0.1\%$ of the qudit's nonlinearity in order for high-fidelity single-qudit gates to be useful in the presence of occupied spectator modes.

\end{abstract}

\maketitle

\section{Introduction}

With the demonstrations of qudit control in quantum devices, such as trapped ions \cite{ringbauer2021universal}, photonic processors \cite{chi2022programmable}, and circuit quantum electrodynamics (cQED) systems \cite{romanenko2020three, wu2020high, alam2022quantum, chakram2022multimode, chakram2021seamless}, many computational levels can be successfully manipulated in order to design and execute quantum algorithms \cite{wang2020qudits}. Compared to its qubit counterparts, high-dimensional quantum computing has many advantages, some of which are lower-depth circuits, noise improvement with hardware-efficient solutions \cite{wang2020qudits, gustafson2021prospects, gustafson2022noise, otten2021impacts} and efficient means for large-scale quantum information experiments to be performed in the lab, such as black hole dynamics modeled as a scrambling unitary  \cite{blok2021quantum}. 

Quantum devices can be controlled optimally via external fields \cite{ma2021quantum, blais2021circuit, koch2022quantum}. Gates can be designed in modules (1- and 2-qudit gates), such as in Ref. \cite{ozguler2022numerical} for bosonic modes. To be able to use synthesized gates in the entire space by preserving their fidelity, one needs to check if the modules function across the entire space. We leverage \texttt{Juqbox.jl} \cite{Juqbox_Github} to synthesize qudit SWAP gates with B-spline parametrization following the techniques in \cite{petersson2020discrete, anders2022optimal}. SWAP operations provide simple, yet effective demonstrations for the effects of frequency shifts, which alter the ideal transitions between energy levels and cause fidelity decay.

We outline the rest of the paper. 
In Section \ref{sec:Sec2}, we provide the effective Hamiltonian of the driven qudit when it interacts with spectator modes, each of which are in Fock states.
In Section \ref{sec:scaling}, the infidelity scaling is given analytically and compared with the numerical result. Finally, in Section \ref{sec:Conclusions}, we conclude the paper by discussing future work, including ways to alleviate the fidelity decay. 

\section{Effective Hamiltonian and frequency shift in the presence of spectator modes}
\label{sec:Sec2}

We focus here on a cQED system with many oscillators/modes. The system Hamiltonian in the rotating frame for each oscillator is given by \cite{ma2021quantum, blais2021circuit}:
\begin{equation}
H = -\sum_{i} \frac{\xi_i}{2} (\hat{n}_i\hat{n}_i -\hat{n}_i) - \sum_{j>i} \xi_{ij}\hat{n}_i \hat{n}_j,
\end{equation}
where $\xi_i$ is the self-Kerr for each oscillator $i$, and $\xi_{ij}$ is the cross-Kerr between oscillators $i$ and $j$. 
If we take the state at time $t$ to be a product state of the form $\ket{\psi}\otimes\ket{\prod_j n_j}$, with the state on the target oscillator $\ket{\psi}$ and spectator modes in Fock states $\{n_j\}$, it is easy to see that the action of the system Hamiltonian will be 
\begin{equation}
\begin{split}
H \ket{\psi} \otimes_j \ket{n_j} &= \Bigl[-\frac{\xi_1}{2} (\hat{n}_1 \hat{n}_1 - \hat{n}_1)\\
       & \quad - \sum_{j>1}  \xi_{1j}n_j\hat{n}_1 + C\Bigr]\ket{\psi}\otimes_j\ket{\prod_j n_j}
       \end{split}
\end{equation}
where $C$ is a constant formed by the action of $H$ on the spectator modes which we ignore from here on as it only generates a global phase. 

The above Hamiltonian does not generate any evolution on the spectator modes (because their initial state is a Fock state), and so focusing only on the target mode and suppressing the subscript $1$ for ease of notation we get the effective Hamiltonian on the target mode:
\begin{equation}
H_{eff} = -\frac\xi2(\hat{n}\hat{n}-\hat{n})-\sum_j \xi_j n_j \hat{n},
\end{equation}
where the first term $H_0 \equiv -\frac\xi2(\hat{n}\hat{n}-\hat{n})$ is the time-independent part of the driven qudit Hamiltonian, $\varepsilon \equiv \sum_j \xi_j n_j$ is the perturbation parameter and $V \equiv -\hat{n}$ is the shift operator appearing due to spectators. In essence, the cross-Kerr between the target mode and the spectator modes in Fock states produces a frequency shift on the target mode. We can write in short:
\begin{equation}
H_{eff} = H_0 + \varepsilon V.
\end{equation}

\begin{figure*}[!htbp]
\includegraphics[scale=0.85, left]{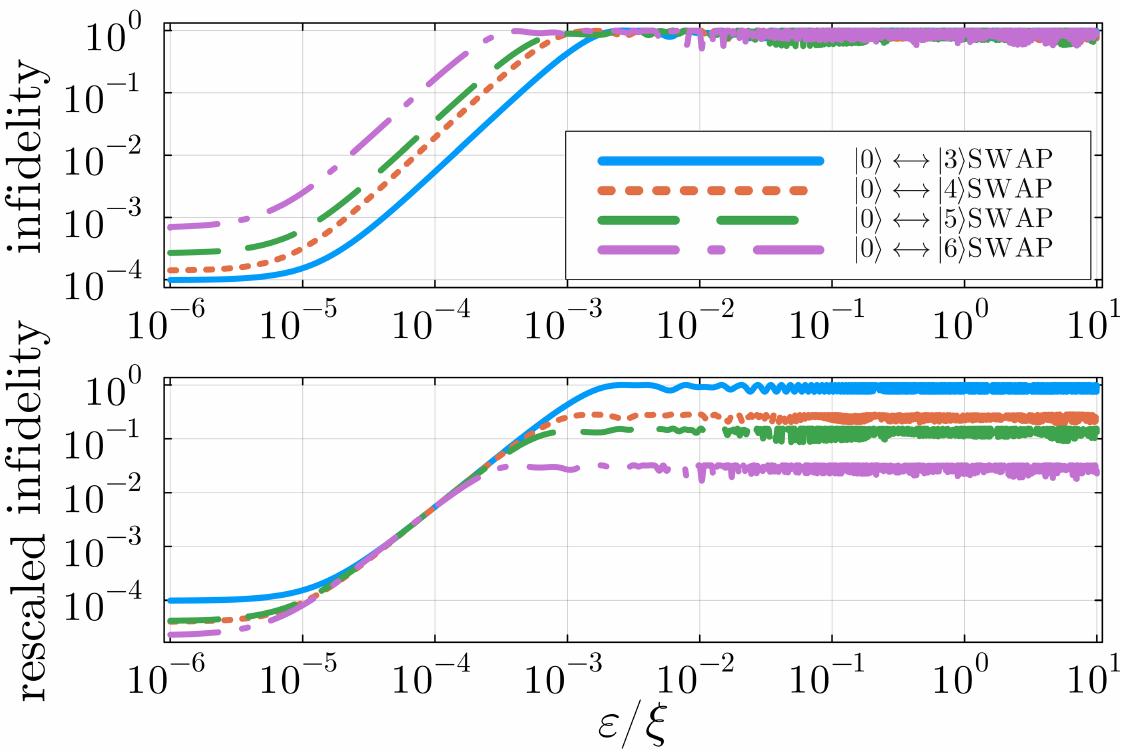}
    \caption{\textbf{(Top)} Infidelity for the labeled SWAP operation arising from a frequency shift $\varepsilon=\sum_j \xi_j n_j$ from the presence of $n_j$ photons in the $j$th spectator mode with cross-Kerr strength to the target mode $\xi_j$ relative to the ideal/target gate. We exclude zero occupation as the x-axis value would be $0$, but the infidelity for that case is that of the optimal control pulse without spectator modes, namely $\order{10^{-4}}$ to $\order{10^{-3}}$ for each gate, as seen at the smallest $\varepsilon$. The slope for small $\varepsilon/\xi=\order{10^{-4}}$ is $\approx 2$, meaning that infidelity scales quadratically with $\varepsilon$. The flat region at very small $\varepsilon$ is the region when the perturbation is negligible, while for larger $\varepsilon$ higher order terms (in part due to saturation near infidelity $\approx 1$) take effect. \textbf{(Bottom)} Rescaled fidelity curves such that the value at the $\varepsilon=10^{-4}$ for each curve is equal, so as to highlight the similarity of the slope in that region.}
    \label{fig:infidelities}
\end{figure*}

\section{Scaling of the infidelity} 
\label{sec:scaling}

For the quantum control problem of gate synthesis, the target action $U$ is known, and we wish to find a time-dependent drive Hamiltonian $H_d(t)$ such that the evolution from $H(t) = H_0 + H_d(t)$ produces the target unitary. The propagator for the driven qudit without spectator modes is given by:
\begin{eqnarray}
    U_0(t) = \mathcal{T} \exp\left[-\frac{i}{\hbar}\int_0^t dt'\left( H_0 + H_{d}(t')\right)\right],
    \label{eq:schroevol}
\end{eqnarray}
where $H_{d}$ is the drive term synthesizing the target gate, i.e., $U_0(t=T)$ is the target gate $U$, $T$ is the gate time and $\mathcal{T}$ is the time-ordering operator. To design a gate using optimal control techniques, we optimize coefficients of pulse control terms $\hat{a}+\hat{a}^\dagger$ and $\hat{a}-\hat{a}^\dagger$ acting on the target oscillator. Thus, they commute with the spectator modes. Using the same drive terms, the effective propagator of the qudit due to spectator shifts is given by:
\begin{eqnarray}
    U_{eff}(t) = \mathcal{T} \exp\left[-\frac{i}{\hbar}\int_0^t dt'\left( H_{eff} + H_{d}(t')\right)\right].
\end{eqnarray}
Here, we assume that the drive field is not coupled to the spectator modes, which is a safe assumption for sufficiently detuned frequencies.

Fidelity between the ideal gate and shifted gate is defined as:
\begin{equation}
F \equiv \Bigg|\frac{Tr(U_{log}(T))}{d}\Bigg|^2,
\end{equation}
where $U_{log}(t) \equiv U_0^\dagger(t) \, U_{eff}(t)$ is the propagator in the \textit{logical frame} of the driven qudit~\footnote{For an in-depth discussion on the concepts of the logical frame and the echo operator, please refer to Eqs. (294-295) in Ref. [21], which corresponds to Eqs. (8-9) in this manuscript.}, $d$ is the norm of $U_0$.

$U_{log}$ is defined in terms of the perturbation as:
\begin{eqnarray}
    U_{log}(t) = \mathcal{T} \exp\left[-\frac{i}{\hbar}\int_0^t dt' \tilde{V}(t') \right],
\label{eq:Ulog_defn}
\end{eqnarray}
where $\tilde{V}(t) \equiv U_0^\dagger(t) V U_0(t)$.
For small perturbation $\varepsilon$, $U_{log}$ is expanded via Baker–Campbell–Hausdorff formula as in Ref. \cite{gorin2006dynamics}:
\begin{eqnarray}
    U_{log}(t) \simeq\ \exp\left[-\frac{i}{\hbar} \Bigl(\varepsilon \bar{V} t + \frac{1}{2}\varepsilon^2 \Gamma(t) + \mathcal{O}(\varepsilon^3) \Bigr)\right],
\label{eq:Ulog_BCH}
\end{eqnarray}
where $\bar{V}$ is the time average of $\tilde{V}(t)$:
\begin{equation}
\bar{V}(t) = \frac{1}{T} \int_0^t \tilde{V}(t') dt',
\end{equation}
and $\Gamma(t)$ is the integral of the time correlation function:
\begin{equation}
\Gamma(t) = \frac{i}{\hbar} \int_0^t dt' \int_{t'}^t dt'' [\tilde{V}(t'), \tilde{V}(t'')].
\end{equation}

$U_{log}(t)$ is:
\begin{equation}
U_{log}(t) \simeq\ I + X + \frac{X^2}{2} + ...,
\end{equation}
where $I$ is the identity matrix and $X \equiv -\frac{i}{\hbar} \Bigl(\varepsilon \bar{V} t + \frac{1}{2}\varepsilon^2 \Gamma(t) \Bigr)$. In simple terms, Equations~\ref{eq:Ulog_defn} and~\ref{eq:Ulog_BCH}, describe the concept of \textit{dynamical decoupling}. They could be used to manipulate the reference frame for a quantum system to isolate and control specific types of interactions, with the goal of preserving the system's quantum state against disturbances, which is crucial for the functioning of quantum computers and error correction in quantum information processing.

Fidelity is then expressed as (suppressing the time parameter of $\bar{V}(t=T)$ for notational simplicity):
\begin{equation}
F \simeq\ 1 - \frac{\Bigl[Tr(\bar{V}^2) - Tr^2(\bar{V})\Bigr] T^2}{\hbar^2 \, d^2} \varepsilon^2.
\label{eq:Fidelity}
\end{equation}
The normalized trace term in the coefficient of $\varepsilon^2$ is the variance of $\bar{V}$ for the maximally mixed state and the coefficient of $\varepsilon^2$ is known as fidelity susceptibility for time-independent systems \cite{gu2010fidelity, ozguler2020response}. We leave the detailed examination of this trace term with time-averaged operators for future work. We compare this analytical scaling ($\sim \varepsilon^2$) with numerical results below ($cf.$ Fig. \ref{fig:infidelities}).

Transitions between the Fock states $\ket{i}$ and $\ket{j}$ in the oscillator are generated by control pulses at the transition frequency between the states, in our case that is $\frac{\xi}{2}(i^2-j^2+i-j)$. The spectator modes, however, shift these frequencies by $ \sum_k \xi_k n_k (i-j)$. To demonstrate the effect of this frequency shift, we can optimize a set of control pulses to produce a swap gate between $\ket{0}$ and $\ket{j}$ on a spectator mode. For concreteness, we use $\omega/2\pi=4.8$ GHz and $\xi/2\pi=0.22$ GHz, with the self-Kerr of the spectator modes being modulated as some fraction of $\xi$ and cross-Kerr parameters equal to $\beta_j \, \xi$ with parameter $\beta_j$ varying for each mode $j$. We use these system parameters so that our gates are directly comparable to those in Section 7 of Ref. \cite{petersson2020discrete}. Other parameters, such as which SWAPs will be generated (SWAPs from $\ket{0}$ to $\ket{3},\ket{4},\ket{5},\ket{6}$) and the time for each gate ($140, 215, 265,$ and $425$ ns respectively) are also taken from that section, along with the use of a single guard level (which implies that a SWAP to state $\ket{k}$ has $k+1$ levels actively participating in the gate and $k+2$ states simulated in the optimization and frequency-shifted calculations). Our only difference is that we restricted our optimization of the control parameters for the ideal (without spectator modes) case  to 200 iterations. Note that our simulations were performed for closed systems but decoherence is not a bottleneck for this work since pulse durations are much shorter than typical coherence times for cQED systems, such as superconducting qubits and cavities \cite{nguyen2019high, romanenko2020three, place2021new, chakram2021seamless, ozguler2021excitation, chakram2022multimode}.

A SWAP gate between level $\ket{i}$ and $\ket{j}$ is defined as:
\begin{equation}
\text{SWAP}_{\ket{i}\longleftrightarrow\ket{j}} = I+\op{i}{j}+\op{j}{i}-\op{i}{i}-\op{j}{j}.
\end{equation}
SWAP gates are vital for shifting matrix elements around and moving quantum states around lattices of qubits, while partial SWAP operations can generate entanglement and more complicated superpositions. Here, our choice of simple SWAPs between the ground state and various excited states of the single oscillator is meant as only an example to illustrate the effect of spectator mode shift of the target oscillator's transition frequencies.

The action on the system Hamiltonian from the spectator modes is conveyed entirely through the term $\varepsilon\hat{V} = -\hat{n} \sum_j \xi_j n_j$. Instead of plotting the infidelity as a function of the populations in adjacent modes, we instead plot it against the parameter $\varepsilon$ (really, $\varepsilon/\xi$ as this is the primary dynamically relevant parameter) in Fig.~\ref{fig:infidelities} for each of the four SWAP gates tested. Here we exclude the zero spectator mode photon occupation case (as $\varepsilon=0$ in that case and thus cannot be placed on a log-log plot). We will simply note that infidelity is $\order{10^{-4}}$ to $\order{10^{-3}}$ in the zero-noise case for each sample. We see that each SWAP gate tested shows the same scaling with $\varepsilon/\xi$ for $\varepsilon/\xi\ll0.001$, scaling with a slope of $\approx 2$ on a log-log plot, denoting quadratic scaling in $\varepsilon$, just as predicted in Section \ref{sec:scaling}. We note that these plots compare the implemented gate with spectator mode state-dependent frequency shifts to the ideal/target gate, and thus the infidelity is lower-bounded by the infidelity of the noiseless gate (resulting in a saturation behavior at small shifts).

To make this even clearer, we also plot a rescaling of these infidelities in Fig.~\ref{fig:infidelities}, with each curve's y-values rescaled such that at a data point for an intermediate value of $\varepsilon$ ($\varepsilon = 10^{-4}$) each curve has the same y-axis value. All the data lines up nearly perfectly for more than an order of magnitude from just over $10^{-5}$ to around $10^{-3.5}$ and only diverges as $\varepsilon/\xi$ approaches $0.001$. We also provide a table, Table \ref{tab:infidelity_slope}, showing the slope of the infidelity curves on the log-log plot in the region of $10^{-4}$, estimated by taking the slope in the region between $10^{-4.05}$ and $10^{-3.95}$, and all of the slopes are near $2$, ie are approximately quadratic.

\begin{table}[]
    \centering
    \begin{tabular}{c | c}
        SWAP gate & infidelity slope \\
        \hline
        $\ket{0}\longleftrightarrow\ket{3}$  &  1.95\\
        $\ket{0}\longleftrightarrow\ket{4}$  &  1.98\\
        $\ket{0}\longleftrightarrow\ket{5}$  &  1.94\\
        $\ket{0}\longleftrightarrow\ket{6}$  &  1.84\\
    \end{tabular}
    \caption{Slope in the region of $\varepsilon/\xi=10^{-4}$ of the infidelity curves to three significant figures for each SWAP gate tested, found by taking the slope of said curve in the region $10^{-4.05}<\varepsilon/\xi<10^{-3.95}$.}
    \label{tab:infidelity_slope}
\end{table}

Thus, we see both from theory and simulation that the effect of spectator modes on the fidelity of a gate generated by a control pulse produced without taking into account spectator modes' frequency shift on the target mode is approximately quadratic in the magnitude of that frequency shift, and rises rapidly to yield an almost orthogonal gate for shifts on the order of $10^{-3}$ times the qudit nonlinearity.\\

\section{Conclusions} 
\label{sec:Conclusions}

We provided a fidelity decay  formula and simulated qudit gates in the presence of spectator modes in order to compare the estimated scaling and numerical results. The fidelity formula, Eq.~\ref{eq:Fidelity}, is independent of the gate, so we expect to get a similar scaling for $\varepsilon \rightarrow 0$ for gates other than SWAP. In our study, a "useful" gate (as mentioned in the abstract) is defined as one that maintains an operational fidelity exceeding 99.9\%, a threshold crucial for fault-tolerant quantum computing. This high fidelity standard ensures the gates are sufficiently reliable for practical quantum computing applications, where precision and error minimization are essential.

We highlight that these frequency shifts yield extremely stringent bounds on interaction parameters and spectator mode occupations. For future directions, one may try to tackle alleviating the effects of fidelity decay with several useful approaches from quantum computing and error correction, such as dynamical decoupling \cite{gorin2006dynamics, lidar2012review, pokharel2022demonstration}, shortcuts to adiabaticity and steering \cite{ozguler2018steering}, circuit optimization and machine learning~\cite{xu2022neural, ogunkoya2024qutrit}, risk-neutral approaches in robust control \cite{anders2022optimal} and bosonic error correction \cite{cai2021bosonic}.

\section*{Acknowledgment}

We thank Jens Koch and Yuri Alexeev for discussions on large-scale simulations. This  material  is  based  upon work supported  by  the U.S. Department of Energy,  Office of Science,  National Quantum  Information  Science  Research  Centers,   Superconducting  Quantum Materials  and  Systems  Center (SQMS) under contract number DE-AC02-07CH11359. We gratefully acknowledge the computing resources provided on Bebop, a high-performance computing cluster operated by the Laboratory Computing Resource Center at Argonne National Laboratory.

\bibliography{refs.bib}

\end{document}